\documentclass[aps,prx,reprint,twocolumn,floatfix,nofootinbib,superscriptaddress,longbibliography]{revtex4-2}
\usepackage{epsfig,amsmath,amssymb,color,comment,physics, dsfont, bbold}
\usepackage[makeroom]{cancel}
\usepackage[caption=false]{subfig}
\usepackage{mathrsfs}
\usepackage[countmax]{subfloat}
\usepackage[normalem]{ulem}
\usepackage[english]{babel}
\usepackage{dsfont}
\usepackage{float}
\usepackage{color}
\usepackage[dvipsnames]{xcolor}
\usepackage{tikz}

\usetikzlibrary{quantikz}

\usepackage[bookmarks=true,colorlinks,linkcolor=blue,urlcolor=NavyBlue,citecolor=RoyalBlue]{hyperref}

\begin{document}

\title{Quantum teleportation between simulated binary black holes}
\author{Aiden Daniel}
\affiliation{School of Physics and Astronomy, University of Leeds, Leeds LS2 9JT, UK}

\author{Tanmay Bhore}
\affiliation{School of Physics and Astronomy, University of Leeds, Leeds LS2 9JT, UK}

\author{Jiannis K. Pachos}
\affiliation{School of Physics and Astronomy, University of Leeds, Leeds LS2 9JT, UK}

\author{Chang Liu}
\affiliation{Shanghai Qi Zhi Institute,  Shanghai 200232, China}
\affiliation{Shanghai Center for Complex Physics, School of Physics and Astronomy, Shanghai Jiao Tong University, Shanghai 200240, China}

\author{Andrew Hallam}
\affiliation{School of Physics and Astronomy, University of Leeds, Leeds LS2 9JT, UK}

\date{\today}
\begin{abstract}

The quantum description of a black hole predicts that quantum information hidden behind the event horizon can be teleported outside almost instantaneously. In this work, we demonstrate that a chiral spin-chain model, which naturally simulates a binary black hole system, can realise this teleportation process. Our system captures two essential components of this protocol: Hawking radiation, which generates the necessary entanglement between the black holes, and optimal scrambling, which enables high-fidelity teleportation on short timescales. Through numerical simulations, we quantify the key timescales governing the process, including the Page time, radiation time, scrambling time, and butterfly velocity, showing their universal dependence on the chiral coupling strength. Our results establish the feasibility of simulating quantum properties of black holes within condensed matter systems, offering an experimentally accessible platform for probing otherwise inaccessible high-energy phenomena.

\end{abstract}
\maketitle

\section{Introduction}\label{sec:introduction}

Quantum teleportation is a fundamental protocol in quantum information theory, enabling the transfer of an unknown qubit state $\ket{\psi}$ from Alice to Bob, provided they share a maximally entangled Bell pair state, $\ket{epr}$~\cite{Bennett1993,Bouwmeester_1997,Boschi1998}. In the standard protocol, Alice performs a joint measurement on $\ket{\psi}$ and her share of $\ket{epr}$, then transmits the measurement outcome to Bob via a classical communication channel. Upon receiving this information, Bob applies a corrective unitary to recover $\ket{\psi}$ with unit fidelity and probability. Despite its simplicity, quantum teleportation has become a cornerstone of modern quantum communication, with experimental demonstrations extending up to 1400 km \cite{Ren2017,takesue2015,sun2017,valivarthi2020,hu2023}.

Beyond quantum communication, teleportation provides a new perspective towards understanding quantum gravity and the black hole information problem \cite{Hawking1976,Strominger_1996}. Hayden and Preskill proposed a black hole adaptation of teleportation, where Alice’s quantum state is thrown into a black hole, and Bob attempts to recover it from outgoing Hawking radiation \cite{Hayden_2007,yoshida2017,Yoshida2019,Landsman2019,Sepehr_2023,Brown_2023}. Unlike standard teleportation, this process bypasses classical communication, instead relying on two key quantum phenomena of black holes: Hawking radiation across the event horizon, which establishes entanglement between the black hole and its exterior, and quantum-chaotic dynamics that generates optimal scrambling deep inside the black holes, which disperses quantum information across the system as rapidly as physically possible~\cite{Sekino_2008}. When the black hole has passed the Page time—the moment when its interior and exterior are maximally entangled—Hawking radiation can be distilled to reconstruct entangled pairs, allowing Bob to probabilistically recover Alice’s state. Both a deterministic and probabilistic approach exists for implementing the teleportation protocol however the latter is slow for exponentially large Hilbert spaces so the former is more commonly used.~\cite{yoshida2017}.

The Hayden-Preskill teleportation protocol is one of the most striking predictions of quantum black hole physics, demonstrating that black holes can act as near-instantaneous mirrors of quantum information. This challenges the classical notion that information falling into a black hole is lost, reinforcing the idea that black holes may function as highly efficient quantum processors. While this phenomenon depends upon the rapid scrambling expected in quantum gravity, a complete theoretical description remains elusive. The Hayden-Preskill protocol is also a fundamentally many-body extension of the traditional few-qubit quantum teleportation. Consequently, various quantum information models and condensed matter analogues have been developed to explore its dynamics \cite{Shenker_2014,Agarwal2023}. 

In this work, we implement the Hayden–Preskill protocol using a chiral spin-chain simulator~\cite{horner2023,forbes2024}. This one-dimensional, strongly correlated system is an analogue black hole model that exhibits geometric features such as Hawking-like radiation with optimal scrambling dynamics, a combination that has remained an open challenge in the field \cite{Banerjee2017,Gu2019}.
Analogue descriptions of the geometric features of gravity have a long history \cite{Barcel_2005,Almeida_2023,Unruh1981,visser1993}, with systems such as superfluids \cite{volovik1995,Kopnin_1998,volovik2000}, Bose-Einstein condensates \cite{Garay_2000,Garay_2001,Kurita_2009} and surface waves in shallow water \cite{Schutzhold_2002, Rousseaux_2008,Rousseaux_2010,Weinfurtner_2011} amongst the experimental platforms that have been proposed. Moreover, optimal scrambling models such as the Sachdev-Ye-Kitaev (SYK) model have been the focus of intensive investigation \cite{Sachdev1993,kitaevtalk2016,Maldacena_2016,Standford2016,Gu2017,Kitaev_2018}. 
Nevertheless, a unified framework capable of capturing both geometry and scrambling has not been achieved so far.
Our approach builds upon and extends a growing body of theoretical and experimental work aiming to simulate black hole thermodynamics and information dynamics using quantum platforms such as cold atoms, trapped ions, and superconducting qubits~\cite{Blok_2021,Landsman2019,Xu2020,Mi_2021,Xu_2019,Xu2024, Shi2023-mx, Kolobov2021-tz,Tian_2022}. By leveraging the unique properties of the chiral spin chain, we provide a complementary and experimentally realistic pathway for probing both semiclassical geometric effects and deeply quantum information-theoretic phenomena within a single, controllable laboratory system.

The remainder of the paper is organised as follows. In Sec.~\ref{sec:model} we introduce the chiral spin-chain model and describe its connection with black hole physics. In Sec.~\ref{black_hole_teleportation}, \ref{chiral_spin_chain_encoding} we explain the Hayden-Preskill protocol and demonstrate its realisation with the chiral spin-chain model. In Sec.~\ref{sec:pagetime}, 
we explore the phenomena necessary to realise the protocol: Hawking radiation and optimal scrambling of information, extending previous exact diagonalisation results on optimal scrambling~\cite{Daniel2024}, by employing a Krylov-subspace method. We also calculate the butterfly velocity of the system and discuss its connection to the success of the teleportation protocol in Methods \ref{methods} . Finally, we summarise our results and discuss their implications in Sec.~\ref{sec:conclusion}. Our results highlight the chiral spin-chain as a powerful platform for studying the interplay of geometric and strong correlation properties of black holes, providing new insights into quantum chaos, the black hole information paradox, and the role of scrambling in quantum teleportation.

\section{Results}

\subsection{Chiral spin-chain simulator of black holes}
\label{sec:model}

\begin{figure}
    \centering   \includegraphics[width=1\linewidth]{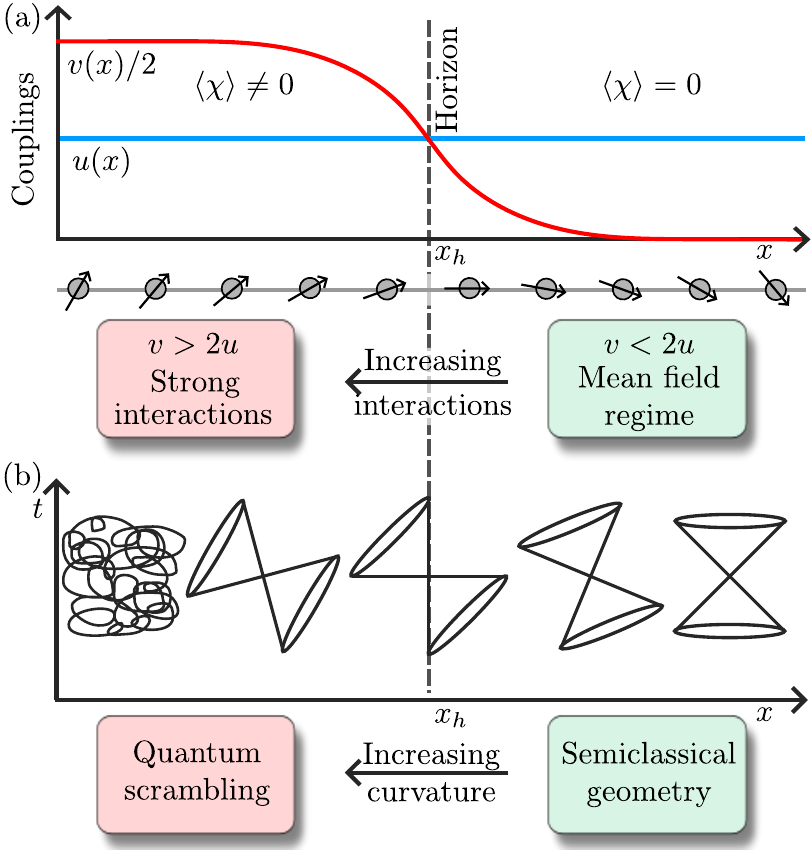}
    \caption{(a) The spin-$1/2$ chiral spin-chain~\eqref{ham}, acts as a quantum simulator of black holes. We take constant coupling $u = 1$ and a spatially varying $v(x) = \alpha \bigl(1 - \tanh[\beta(x - x_h)]\bigr)$. This profile creates two distinct regions: a strongly interacting chiral region ($\langle \chi \rangle \neq 0$) for $0 < x < x_h$ where $v > 2u$, and a weakly interacting non-chiral region ($\langle \chi \rangle = 0$) for $x > x_h$ where $v < 2u$. In the weak-fluctuation regime ($v \lesssim 2u$), mean-field theory applies and the continuum limit corresponds to Dirac fermions propagating on a curved black hole background described by metric~\eqref{eqn:metric}. In this regime the chaotic behaviour is weak. The horizon lies at $x = x_h$, defined by $v = 2u$. Inside the black hole ($v > 2u$), the MFT description fails and strong interactions give rise to optimal scrambling, as expected in the black hole interior. (b) Schematic of a quantum-field description of a black hole in the Gullstrand-Painlev\'e coordinates. Increasing curvature progressively tilts the lightcones; beyond the horizon, both lightcone directions point inward, preventing classical particles to escape. Deep inside the black hole, large curvature ultimately invalidates the semiclassical geometric picture, leading to strong quantum fluctuations and rapid scrambling of information.
    }
    \label{fig:sketch_chiralmodel}
\end{figure}

We begin by introducing the chiral spin-chain model of spin-$1/2$ particles and outlining its role as an effective simulator of key quantum properties of $(1+1)$D black holes. In particular, we show how carefully engineered coupling configurations can generate a region that is accurately described by a curved background geometry, capturing the semiclassical gravitational effects of the horizon, and a distinct region governed by optimal scrambling dynamics, as expected in a fully quantum regime deep inside the black hole. This unified framework allows us to faithfully reproduce both geometric and chaotic features within a single, controllable platform, as shown in Fig.~\ref{fig:sketch_chiralmodel}.

The model is governed by the Hamiltonian
\begin{equation}
H \!=\! \frac{1}{2} \!\sum_{n=1}^N \!\left[ -u \left( S^x_n S^x_{n+1} \!+\! S^y_n S^y_{n+1} \right)\! +\! \frac{v}{2} \boldsymbol{S}_n \!\cdot\! \left( \boldsymbol{S}_{n+1} \!\times\! \boldsymbol{S}_{n+2} \right) \right],
\label{ham}
\end{equation}
where $u$ and $v$ are, in general, position-dependent real functions. Here, $\boldsymbol{S}_n = (\sigma^x_n/2, \sigma^y_n/2, \sigma^z_n/2)$, with $\sigma_n^\alpha$ ($\alpha = x,y,z$) denoting the Pauli matrices at site $n$.

This Hamiltonian extends the standard XY model by incorporating a three-spin chirality term $\chi_n = \boldsymbol{S}_n \cdot (\boldsymbol{S}_{n+1} \times \boldsymbol{S}_{n+2})$, which introduces non-integrable interactions and enriches the model's dynamical behaviour~\cite{Pachos05,Tsomokos2008,horner2023,forbes2024}. Through a Jordan-Wigner transformation, Hamiltonian~\eqref{ham} can be mapped to a model of interacting fermions, further highlighting its strongly correlated nature.

For constant couplings $u$ and $v$, Hamiltonian~\eqref{ham} exhibits two distinct phases. When $|v|$ is small compared to $|u|$, the ground state expectation value of the chirality, $\langle \chi \rangle = \langle \sum_n \chi_n \rangle$, vanishes. In this regime, the interactions are weak and can be treated perturbatively, allowing the system to be accurately described by the mean-field theory (MFT) approximation as a free fermionic model. In the continuum limit, this free fermion description corresponds to Dirac fermions in a curved spacetime background, as we discuss below.

Conversely, when $|v|$ becomes large compared to $|u|$, the system enters a strongly interacting phase where $\langle \chi \rangle \neq 0$ and the MFT description breaks down. In this regime, the dynamics become chaotic, and quantum information rapidly spreads throughout the system when evolved under Hamiltonian~\eqref{ham}. Notably, it has been shown in~\cite{Daniel2024} that for $|v| \gg |u|$, Hamiltonian~\eqref{ham} exhibits optimal scrambling, saturating the upper bound on information delocalisation,  similarly to quantum black holes and their holographic duals, such as the SYK model~\cite{Standford2016,Kobrin2021}.

We can describe the geometric behaviour of Hamiltonian~\eqref{ham} using a mean-field theory (MFT) approximation, which is valid when quantum correlations are weak and fluctuations are small. This approach parallels the semiclassical approximation in quantum gravity, where quantum fields evolve freely on a fixed curved spacetime background. Although semiclassical gravity is conventionally formulated using bosonic fields, in $(1+1)$ dimensions the low-energy behaviour of free Dirac fermions is mathematically equivalent to that of free bosonic fields \cite{senechal1999, Miranda}.

Within this MFT framework, the fermionic Hamiltonian simplifies to
\begin{equation}
H_{MFT} = \sum_{n=1}^N \left(uc_n^\dagger c_{n+1} + \frac{i v}{4} c_n^\dagger c_{n+2}\right) +\rm{h.c.}
\label{eq:MFT}
\end{equation}
To analyse $H_{MFT}$, we consider homogeneous $v$, introduce a two-site unit cell with sublattices $a$ and $b$, and impose periodic boundary conditions to extract the dispersion relation. In the continuum (low-energy) limit, the dispersion relation of $H_{MFT}$ is captured by the Dirac action on a curved spacetime background~\cite{horner2023},
\begin{equation}
S_{\rm{MFT}} = \int \mathrm{d}^{1+1}x \, |e| \, \bar{\psi}(x)\left( ie_a^{\ \mu} \gamma^a \overset{\leftrightarrow}{\partial_\mu} \right) \psi(x),
\label{eq:cont_act}
\end{equation}
where $a=0,1$ are local Lorentz indices and $\mu = t,x$ are spacetime coordinates. The spinor field is defined as $\psi(x) = (a(x), b(x))^T / \sqrt{|e|}$, with $A \overset{\leftrightarrow}{\partial_\mu} B = \frac{1}{2}(A \partial_\mu B - (\partial_\mu A)B)$ and $\gamma^a = (\sigma^z, -i\sigma^x)$. The determinant $|e| = \det(e^a_{\ \mu})$, where the zweibein $e_a^{\ \mu}$ relates to the spacetime metric via $g_{\mu\nu} = e^a_{\ \mu} e^b_{\ \nu}\eta_{ab}$, with $\eta_{ab} = \mathrm{diag}(1,-1)$ the Minkowski metric. The resulting metric reads
\begin{equation}
\mathrm{d}s^2 = \left( 1 - \frac{v^2}{4u^2} \right)\mathrm{d}t^2 - \frac{4v}{u^2}\mathrm{d}t \mathrm{d}x - \frac{16}{u^2}\mathrm{d}x^2.
\label{eqn:metric}
\end{equation}
This corresponds to the Schwarzschild black hole metric expressed in Gullstrand-Painlevé coordinates~\cite{Volovik2003,Volovik1,Volovik2,Volovik3}, which are regular at the horizon and natural from the viewpoint of an infalling observer. As $v$ transitions from $|v| < 2|u|$ (outside the black hole) to $|v| > 2|u|$ (inside), the sign of the time component flips at $|v| = 2|u|$, tipping the lightcones so that all future-directed paths point inward, preventing escape. This geometric description faithfully reproduces the phase diagram of the MFT Hamiltonian~\eqref{eq:MFT}, with a critical point at $|v| = 2|u|$ separating the non-chiral and chiral regimes.

The above discussion focuses on constant coupling parameters. To emulate black hole geometry in our simulator, we instead keep $u$ constant and allow $v(x)$ to vary smoothly in space, ensuring that the continuum approximation described by Eqs.\eqref{eq:cont_act} and \eqref{eqn:metric} remains valid. This spatial modulation of the chiral coupling generates an interface between chiral and non-chiral phases precisely at $|v| = 2|u|$, which can be interpreted as the event horizon of a black hole. For concreteness, we set $u=1$ and choose $v(x) = \alpha\big(1 - \tanh[\beta(x - x_h)]\big)$, where $x_h$ specifies the horizon position, as depicted in Fig.~\ref{fig:sketch_chiralmodel}(b). Although this coupling profile does not strictly satisfy the Einstein equations, it effectively captures the essential geometric and dynamical features of black hole physics required for our study.

The MFT description is valid in the regime where $|v| \lesssim 2|u|$, corresponding to relatively weak interactions and suppressed quantum fluctuations. This parameter range describes the spatial region near and outside the horizon, $x_h \lesssim x$, which can be effectively modelled numerically even for large system sizes. In this regime, the simulator successfully captures semiclassical phenomena such as Hawking radiation~\cite{horner2023,forbes2024}. By contrast, in the region with $|v| > 2|u|$, i.e., $x < x_h$, the system enters a strongly interacting phase where the MFT breaks down. Here, exact diagonalisation studies for smaller systems have revealed that the chiral spin-chain exhibits pronounced quantum chaotic dynamics~\cite{Daniel2024}. Unlike classical systems, where chaotic evolution can be unboundedly fast, quantum dynamics are constrained by unitarity~\cite{Maldacena_2016}. Remarkably, Hamiltonian~\eqref{ham} has been shown to saturate this fundamental upper limit, achieving so-called optimal scrambling, a key signature of quantum black holes and their holographic duals such as the SYK model~\cite{Hawking1993,Wald2001,Sachdev1993,Standford2016,kitaevtalk2016,Kitaev_2018,polchinski2016spectrum,Fu2016,Kobrin2021}. This behaviour is probed through the early-time growth of out-of-time-ordered correlators (OTOCs), quantified by the Lyapunov exponent $\lambda$, which approaches the maximal bound $\lambda = 2\pi T$ in fully chaotic systems~\cite{Maldacena_2016}. By contrast, in the weakly interacting regime, one finds a quadratic dependence $\lambda \propto T^2$, which becomes subdominant compared to the linear growth for $T \to 0$~\cite{Daniel2024}.

From the quantum properties of the chiral spin-chain simulator with spatially varying couplings, a novel representation of a quantum black hole emerges. In the region near and outside the horizon ($x \gtrsim x_h$), the simulator is well-described by the mean-field theory, capturing the behaviour of quantum fields propagating on a classical curved spacetime background given by~\eqref{eqn:metric}. This semiclassical regime is responsible for phenomena such as Hawking radiation. In contrast, deep inside the black hole, $x < x_h$, the system transitions into a regime of optimal scrambling, mirroring the chaotic dynamics expected in the black hole interior. These two distinct regions, illustrated schematically in Fig.~\ref{fig:sketch_chiralmodel}(b), are seamlessly unified within a single physical model by tuning the spatial profile of the couplings in Eq.~\eqref{ham}. This integrated framework simultaneously captures the semiclassical geometry and the fully quantum chaotic features, offering a uniquely powerful platform for simulating black hole physics. In particular, it enables the realisation of complex phenomena such as the Hayden–Preskill quantum teleportation protocol, which crucially relies on the interplay between Hawking radiation at the horizon and optimal scrambling in the interior.


\begin{figure}
    \centering   \includegraphics[width=1\linewidth]{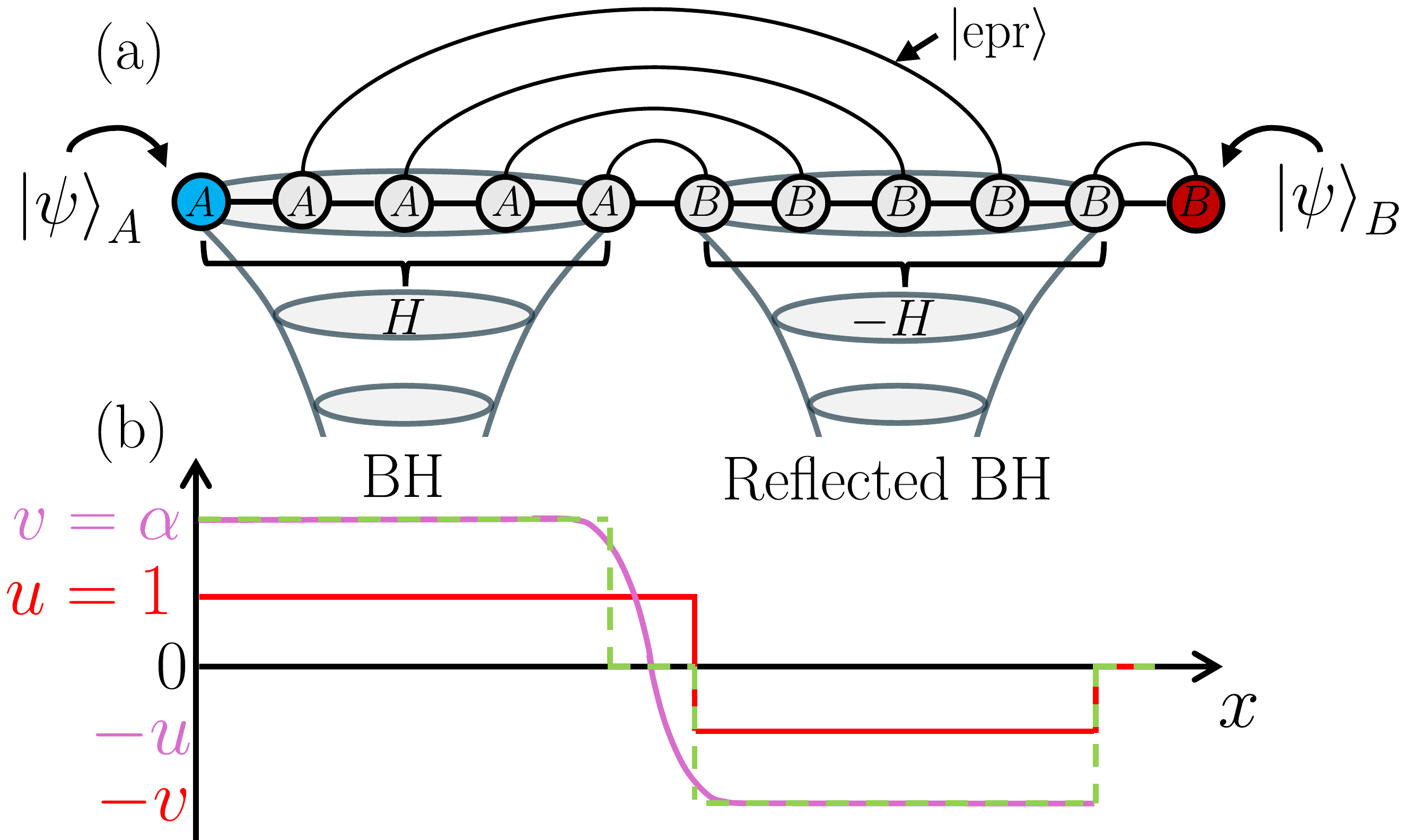}
    \caption{ (a) Schematic of the chiral spin-chain with the encoded state $\ket{\psi}_A$ and the initially prepared EPR states, $\ket{epr}$. To implement the Hayden-Preskill protocol we employ the scrambling Hamiltonian $H$ inside the black hole and $-H$ on its outside, with left and right halves of the chain corresponding geometrically to mirrored black holes. The qubits Alice and Bob initially have access to are labelled A and B respectively. (b) The coupling profile for $u$ (red) and $v$ (green/purple) along the spin-chain for the Hamiltonian in Eq.~\eqref{ham}. Using Eq.~\eqref{eqn:couplings} and taking $\alpha=v$, we highlight the difference in the $v$ coupling between the (smooth purple) geometric  interpretation which takes varying $\beta$ where large free system can be numerically considered and the (dashed green) teleportation protocol in a small strongly interacting system where $\beta=\infty$. To realise the Hamiltonian $H\oplus (-H)$ the two-site $u$ term flips after the site $(N-1)/2$, while the 3-site $v$ term is 0 on site $(N-1)/2$ but becomes $\pm v$ on either side.}
    \label{fig:sketch}
\end{figure}

\subsection{Black hole teleportation}\label{black_hole_teleportation}

Hayden and Preskill proposed a scheme for teleporting the quantum information stored in a black hole. Alice, who resides inside the black hole, seeks to teleport the quantum state $\ket{\psi}_A = \alpha\ket{0} + \beta\ket{1}$ to Bob who is positioned outside the black hole. As classical communication is forbidden due to the event horizon, the teleportation protocol instead relies on quantum correlations established through $L=(N-3)/2$ EPR pairs $\ket{EPR} = \ket{epr}^{\otimes L}$, where $\ket{epr} = \frac{1}{\sqrt{2}}(\ket{01}_{AB} - \ket{10}_{AB})$, arranged in a nested fashion, as shown in Fig.~\ref{fig:sketch}(a), and an optimally scrambling evolution $U$ within the black hole \cite{Sekino_2008,Maldacena_2016,kitaevtalk2016}. Alice's system $A$ now consists of a state $\ket{\psi}_A$ on site $n=1$ and one qubit from each of the $L$ Bell pairs on sites $n=2$ to $n=(N-1)/2$ where $N$ is the system size which we take to be odd. Bob's system is the remaining qubits from the $L$ Bell pairs on sites $n=(N+1)/2$ to $n=N-2$, and a final $\ket{epr}$ between sites $n=N-1$ and $N$. This state can be rewritten as
\begin{eqnarray} 
&&\ket{\psi}_A \otimes \ket{EPR}\otimes\ket{epr}_{N-1,N} = \nonumber \\ 
&&\frac{1}{2}\ket{epr}_{A,N-1} \otimes \ket{EPR} \otimes \ket{\psi}_B \nonumber \\
&&+\frac{1}{2}\!\!\sum_{a={x,y,z}}\!\!\sigma^a_A\ket{epr}_{A,N-1} \otimes \ket{EPR}  \otimes \sigma^a_B\ket{\psi}_B.
\label{eqn:telestate_mb} 
\end{eqnarray}
Crucially, the first part of Eq. \eqref{eqn:telestate_mb}, $\frac{1}{2}\ket{epr}_{A,N-1} \otimes \ket{EPR} \otimes \ket{\psi}_B \nonumber$,  is left invariant by Alice applying $U$ on sites $n=1$ to $n=(N-1)/2$ provided Bob applies a corresponding unitary $U^*$ to sites $n=(N+1)/2$ to $N-1$ at their end. Despite this, such a unitary will in general scramble the remaining part of the state, including the central $L$ EPR pairs. Therefore, if Bob performs $E$ many EPR measurements and post-selects the outcomes to be $\ket{epr}$, with high probability the state will originate from the first part of \eqref{eqn:telestate_mb}, and Bob will have successfully teleported the state. Rapid scrambling of the state is vital to achieve this result, so that the EPR measurements have a low probability of reproducing $\ket{epr}$ for the second part of the state. Assuming full thermalisation of the state, the probability of Bob successfully post-selecting an epr pair $ P_E =\frac{1}{4} + \frac{3}{4^{E+1}}$, and fidelity of the teleported state $F_E =  1 - \frac{2}{(4^E+3)}$, tend to $\frac{1}{4}$ and 1, respectively, exponentially fast as $E$ increases~\cite{Agarwal2023}. The faster the system thermalises, the faster this result is obtained. 

\subsection{Chiral spin-chain encoding}\label{chiral_spin_chain_encoding}

\begin{figure}
    \centering
    \includegraphics[width=1\linewidth]{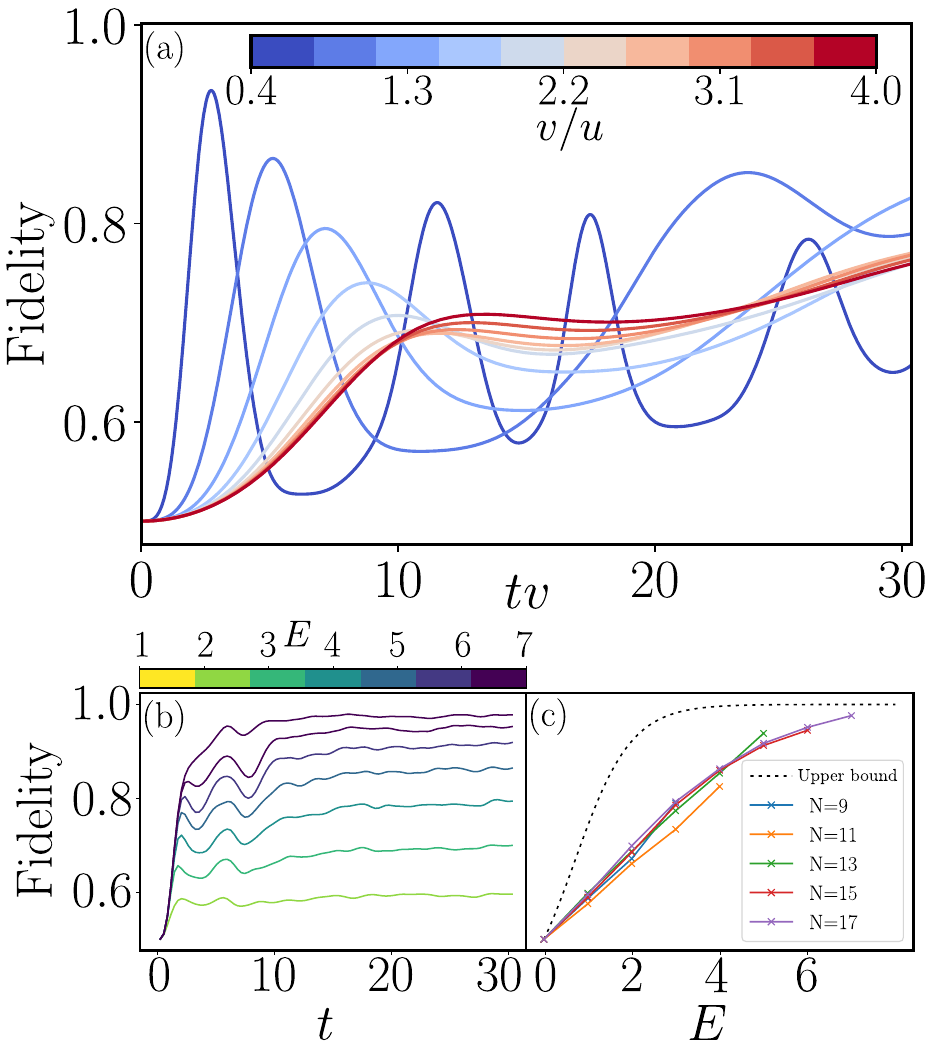}
    \caption{(a) Fidelity of teleportation over time, $t$, scaled by $v$ after quenching the initial state, $\ket{\psi}_A\otimes\ket{EPR}$, with Hamiltonian \eqref{ham}. Here we take $\alpha=v$ and $\beta=\infty$ in Eq.~\eqref{eqn:couplings} with the slight modification such that no $v$ coupling is allowed across the two halves of the chain as shown in Fig.~\ref{fig:sketch}(b). Success is determined by the overlap between the final state $\ket{\psi}_B$ of the last site with the initially prepared state $\ket{\psi}_A$. Colour indicates the value of $v$ during the quench where we see a notable change in teleportation success across the phase transition at $v/u\approx 2$. Four EPR pairs, $\ket{epr}$, are measured in this case. 
    (b) Similarly but unscaled in time and at a fixed $v=8$. Colour corresponds to a different number of EPR pairs measured, $E$, where we see teleportation success increases with $E$ as expected. Here we take $N=17$ and $u=1$ using exact diagonalisation (ED). (c) Taking a slice at $t=30$ in (b) for different system sizes, we see how the scaling of fidelity with $E$ changes for system size $N$ when compared to the theoretical upper bound of $F_E$. We see with increasing system size our results tend towards the optimal theoretical prediction. Code used is publicly available at~\cite{danielgithub}.}
    \label{fig:chiralteleportation}
\end{figure}

To employ the chiral spin-chain simulator, Alice introduces her state $\ket{\psi}$ inside the black hole, where it evolves under the Hamiltonian $H$ given in \eqref{ham} for time $t$, producing the scrambling unitary $U = e^{-iHt}$. The nested structure of the $\ket{epr}$ states allows Bob to apply the counter evolution $MU^*M = e^{iHt}$ outside the black hole. $M$ is a unitary operator which spatially inverts the order of the spins, leaving the $\ket{EPR}$ state invariant, as well as the Hamiltonian since $MH^*M=H$. As a result, the total evolution for the system is given by $e^{-iHt} \otimes e^{iHt} \otimes I_N$, where $I_N$ is the identify acting on the $N$th spin. 

In the chiral model, this evolution is realised by introducing a coupling configuration $H \oplus (-H)$ that corresponds to a binary black hole system. We emphasise that neither $H$ acts upon the final spin of the chain, as shown in Fig.~\ref{fig:sketch}(a). A step function profile for $u(x)$, where $u$ changes sign at the centre of the chain, and a $\tanh$ profile for $v(x)$, are given by
\begin{equation} u(x) = -{\rm sgn}(x-x_0), \,\,\,
v(x) = \alpha \tanh[\beta(x - x_0)] + 1,
\label{eqn:couplings} 
\end{equation}
where $x_0 = aN/2$ is the central site of the spin chain and $a$ is the lattice spacing. The gradient of $v(x)$, and hence the Hawking temperature, is controlled by tuning $\alpha$ and $\beta$. This configuration creates Hamiltonian $H$ in one half of the system and $-H$ in the other. From the metric in \eqref{eqn:metric} we see that the black hole horizons appear at position $x_h$ at which $|v(x_h)|/2 = |u(x_h)|$, while the interiors reside at the chain's edges, where $|v(x)|>2|u(x)|$, with a small shared exterior region at the centre, where $|v(x)|<2|u(x)|$. An additional XY-Hamiltonian at the exterior of the black holes leaves the total $\ket{EPR}$ state unchanged. Throughout this work we couple the two chains with a $u(aN/2)=+1$ (with lattice spacing $a=1$). We also allow non-trivial hopping between the two halves of the system, as shown in Fig.~\ref{fig:sketch}(b), though the results are not sensitive to this choice. Information is encoded in the left black hole by Alice and then the system is evolved for time $t$ before $E$ many EPR measurements are applied and the fidelity of the teleportation, $F$, is monitored. This time, $t$, also allows for the chain to sufficiently radiate out of the first black hole such that Bob can perform the measurements on Alice's original qubits. This process is illustrated through the Page curve discussed later in Section.~\ref{sec:pagetime}.

The time evolution of the fidelity is shown in Fig.~\ref{fig:chiralteleportation}(a). We observe that for small chiral couplings, $v/2<u$, the fidelity oscillates without the ability to produce successful teleportation for a sustained period of time. On the other hand, when $v/2>u$ the fidelity increases and remains high and stable. This stability is seen to increase with system size. Further results discussing the relevance of system size and number of measurements for the teleportation protocol can be found in the Supplementary material. The contrast between oscillatory and constant behaviour is due to the absence or presence of interactions that cause scrambling in the initial $\sigma^a\ket{EPR}$ state. For example, for the non-interacting mean field theory defined in Eq.~\eqref{eq:MFT}, oscillatory behaviour is observed for all values of $v$. When the system is in the strong chiral phase, we observe an increase in the fidelity when the number of measurements is increased. This is shown in Fig.~\ref{fig:chiralteleportation}(b) where the fidelity is plotted against time. In fact, the increase of the teleportation fidelity is exponential with the number $E$ of EPR measurements, as shown in Fig.~\ref{fig:chiralteleportation}(c), guaranteeing a fast convergence to unit fidelity. As a result, the chiral spin-chain can successfully reproduce the Hayden-Preskill teleportation protocol. 

\subsection{Hawking radiation and Page time}\label{sec:pagetime}

\begin{figure}
\includegraphics[width=1\linewidth]{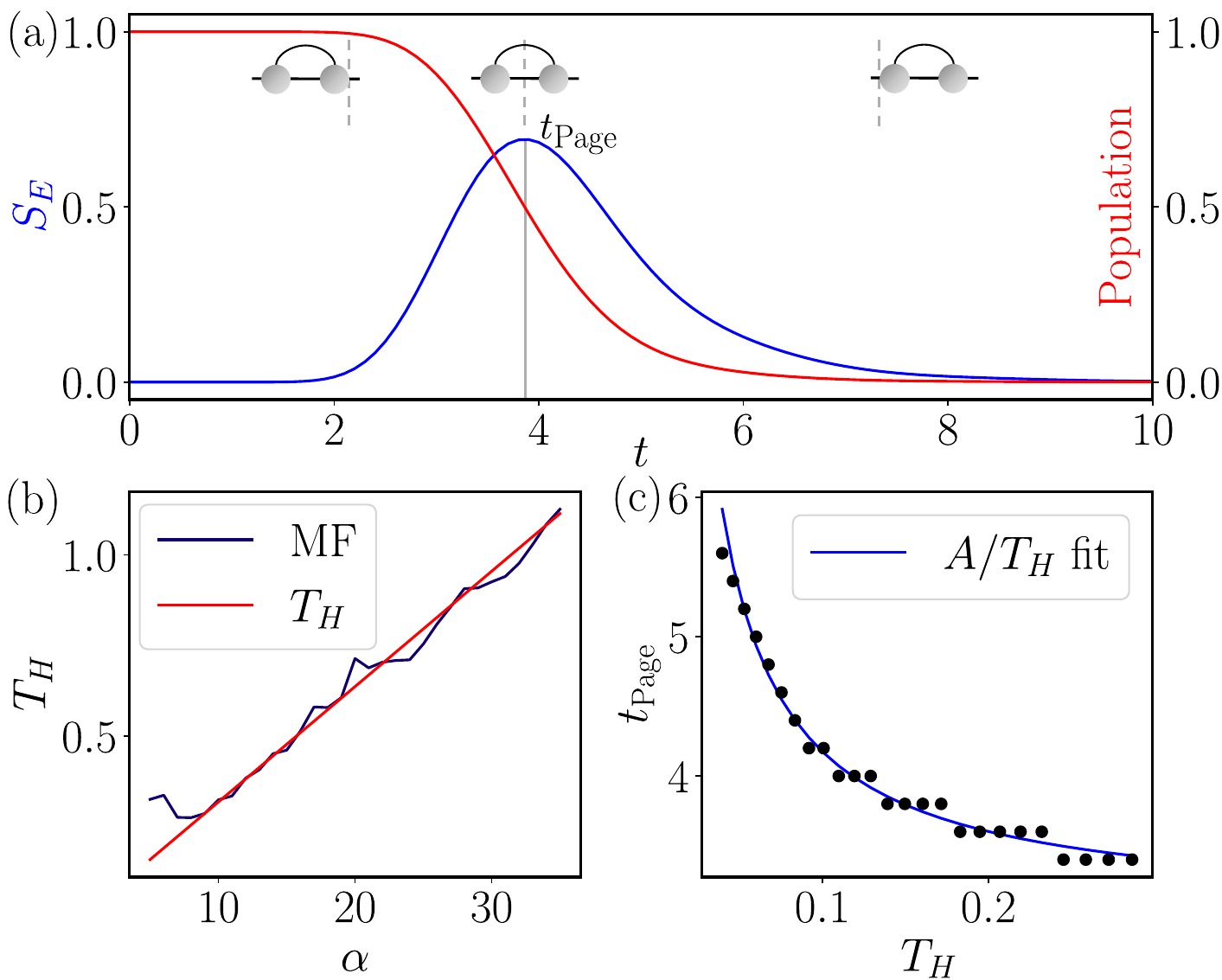}
\caption{Page curve, Page time, and Hawking temperature of the mean-field (MF) chiral Hamiltonian~\eqref{eq:MFT}.  
    (a) The entanglement entropy, $S_E$ (blue line), of the black hole across the horizon as a function of time, $t$, during the Hawking radiation of a single particle. This curve, known as the Page curve, reaches its maximum at the Page time, $t_{\rm{Page}}$, when the population inside the horizon (red line) is halved. To accelerate black hole evaporation, we consider a single-particle population inside the black hole and place the horizon at $n_h=2$, while the entropy is evaluated for a bipartition at $n=22$ to account for the entropy of particles that are completely free from the black hole attraction. The inset sketches illustrate the evolution of an EPR pair as it crosses the event horizon (dashed line): at early times, the EPR pair is entirely inside the black hole ($S_E = 0$); at $t_{\rm{Page}}$, half of the pair escapes, yielding maximal entanglement ($S = \ln 2$); at late times, the second half is also emitted, leading to $S \to 0$.  
    (b) The Hawking temperature $T_H$ as a function of $\alpha$ in the MF model (blue line) compared to its analytical value $T_H = \alpha \beta / 2 \pi$ (red line). Here, $N = 500$ and the horizon is at $n_h = 250$.  
    (c) Page time $t_{\rm{Page}}$ as a function of $T_H$ in the MF model (black dots), alongside the fitted curve $A/T_H$ (blue line) with $A = 0.278$, obtained numerically by varying $\alpha$ and $\beta$.}
\label{fig:Hawking_temp_u_step}
\end{figure}

An essential mechanism in black hole teleportation is Hawking radiation that builds quantum correlations between the inside and the outside of the black hole. This effect can be analysed with the semiclassical limit of the black hole, which corresponds to the mean field theory of the chiral spin-chain model. We demonstrate that Hawking radiation can generate the quantum correlations necessary for teleportation within a binary black hole system. Furthermore, we show that the coupling configuration \eqref{eqn:couplings}, adopted for black hole teleportation, preserves the geometric properties of the metric \eqref{eqn:metric} at the horizon, as encoded in the temperature of the Hawking radiation.


Here, we utilise the mean-field theory (MFT) description of the chiral spin chain given in \eqref{eq:MFT}. We begin by determining the Page curve, which tracks the entanglement entropy of the black hole across the horizon during Hawking radiation, considering a single-particle population inside the black hole, with the exterior initially empty. We also assume a small horizon radius, $n_h = 2$, ensuring that complete tunnelling of the particle across the horizon occurs within a reasonable timescale. Initially, the system is in a tensor product state across the horizon, corresponding to an entanglement entropy of $S_E = 0$. As the particle tunnels outward, the fermionic mode inside the black hole becomes entangled with modes outside, leading to an increase in entanglement. When the population inside the black hole is halved, the entropy reaches its maximum, $S_E = \ln 2$. Subsequently, as the black hole population asymptotically approaches zero, the entanglement entropy decays to $S_E \to 0$.  This simple model reproduces the Page curve shown in Fig.~\ref{fig:Hawking_temp_u_step}(a), with the time at which entropy is maximised defining the Page time, $t_{\rm{Page}}$. A more precise simulation of black hole evaporation would require more particles inside the black hole and a simultaneous reduction in the horizon radius reflecting the loss of mass due to Hawking radiation~\cite{Hawking1971,Bekenstein1972,Bekenstein1973,Bardeen1973,Bekenstein1974,Hawking1974,Hawking1975,Hawking1976_blackholes,Rishabh2025,DegerUnpublished}. Nevertheless, our process convincingly demonstrates how the simulated black hole becomes entangled with its exterior, a key requirement for executing the Hayden-Preskill protocol. Full details of this calculation can be found in the Supplementary Material.

Next, we examine the Hawking radiation temperature, $T_H$. The same configuration is used, a single particle is placed inside the black hole and the coupling pattern given in~\eqref{eqn:couplings} is chosen, defining a binary black hole system. To accurately reproduce the Hawking temperature, we use a large system, $N=500$, with the horizon positioned in the middle, $n_h=N/2$. The particle inside the black hole is initialised at $n = 230$ and evolves under the MFT Hamiltonian~\eqref{eq:MFT}. By varying the curvature at the horizon, e.g., by tuning $\alpha$, the temperature of the outgoing radiation follows the theoretical prediction
\begin{equation}
T_H = \frac{1}{2\pi} \left| \frac{dv(v_h)}{dx} \right| = \frac{\alpha \beta}{2\pi}.
\end{equation}
Previously, it was demonstrated that an analogous system with uniform $u(x)$ yields a consistent $T_H$ across a range of temperatures and initial states \cite{horner2023}.
Although the sign change in $u$, introduced in \eqref{eqn:couplings} for the implementation of the teleportation protocol, leaves the metric \eqref{eqn:metric} unaffected, the derivation of this metric assumes uniform couplings. This raises the question of whether the dynamical properties of the chain, such as Hawking radiation, remain unchanged. To verify this numerically, we extract the Hawking temperature from radiation emitted through the horizon after a dynamical quench inside the black hole. Fig.~\ref{fig:Hawking_temp_u_step}(b) shows the Hawking temperature as a function of $\alpha$, demonstrating excellent agreement with theoretical predictions across a wide range of couplings. This confirms that the coupling configuration \eqref{eqn:couplings} accurately reproduces the black hole geometry, yielding Hawking radiation with temperature $T_H = \frac{\alpha\beta}{2\pi}$.

Modelling the black hole with a chiral spin chain allows us to identify the characteristic timescales governing the Hayden-Preskill teleportation protocol. The teleportation process can only proceed after the Page time, since this ensures that quantum matter inside the black hole is maximally entangled with the exterior. Only beyond this point can Hawking radiation reliably transmit EPR pairs across the horizon, facilitating the Hayden-Preskill protocol.
The Page time, $t_{\text{Page}}$, marks the moment when the entanglement entropy of the emitted Hawking radiation reaches its peak, given by
$t_{\text{Page}} = \frac{S_E}{2\pi T_H}$, where $S$ is the black hole entropy~\cite{Page1993,Page1980}. Subsequent Hawking radiation, which provides additional EPR pairs to Bob, follows the emission timescale
\begin{equation}
t_{\text{rad}} = \frac{\Delta S_E}{2\pi T_H},
\end{equation}
where $\Delta S_E$ represents the corresponding entropy change. Fig.~\ref{fig:Hawking_temp_u_step}(c) confirms that the Page time, obtained by varying $\alpha$ and $\beta$, is inversely proportional to $T_H$. Since $T_H \propto v$, the radiation emission time $t_{\text{rad}}$ scales inversely with the chiral coupling $v$.

\section{Methods}\label{methods}


We now discuss the different physical effects that are involved in the black hole teleportation process aiming to determine the overall duration our quantum simulator takes to perform the Hayden-Preskill protocol such as the geometry encoding and optimal scrambling, identifying the relevant timescales.

\subsection{Optimal scrambling}\label{optimalscrambling}

\begin{figure}
\includegraphics[width=1\linewidth]{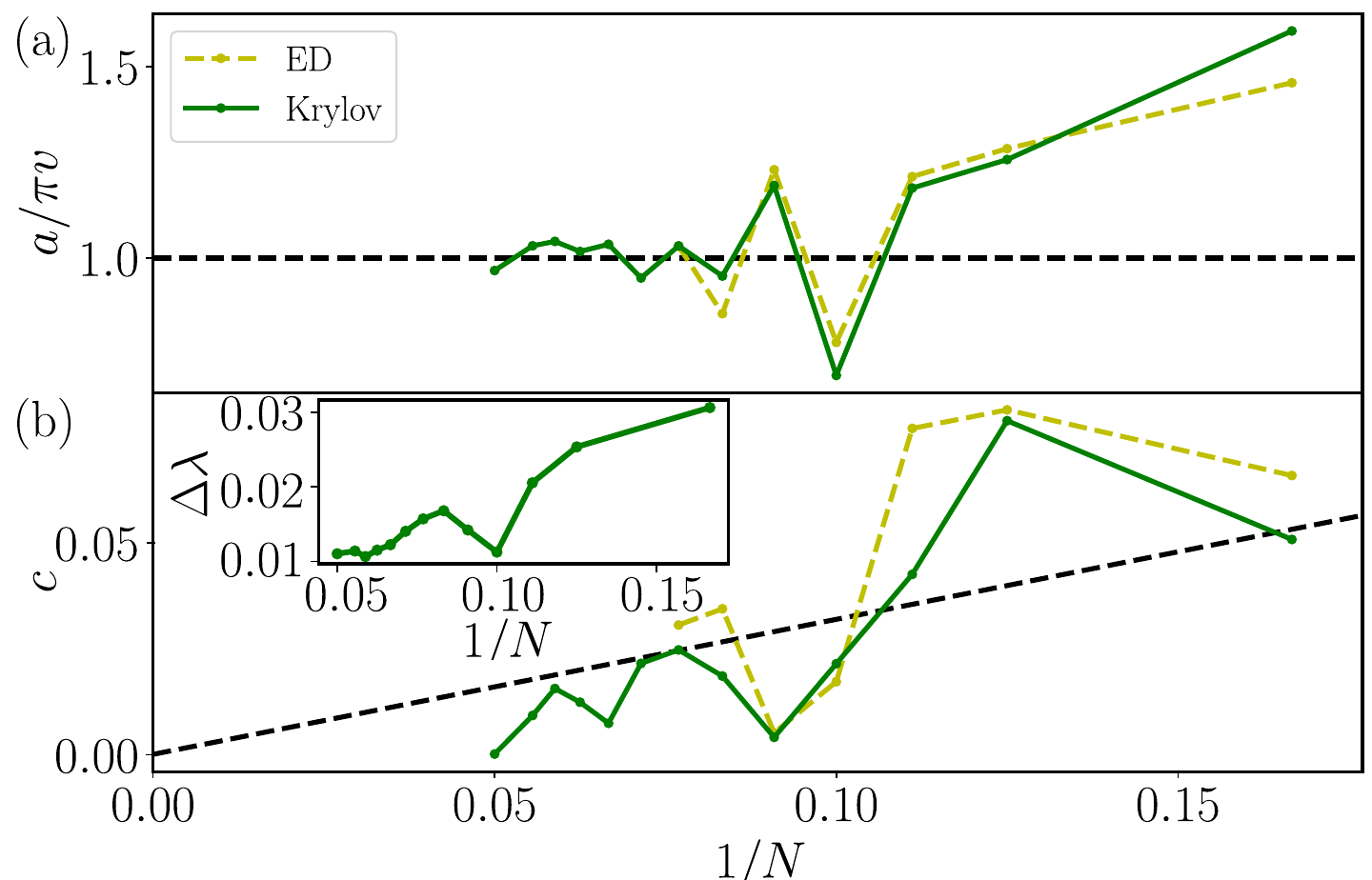}
\caption{ The Lyapunov exponent $\lambda$ of the chiral spin-chain model \eqref{ham} in the strongly interacting regime where $v=8$, $u=1$. For small temperatures $T$ the Lyapunov exponent takes the form $\lambda = aT+c$ for $a$ and $c$ some fitting parameters. As $N$ increases the fitting parameters approach $a \to 2\pi \frac{v}{2}$ and $c \to 0$, thus recovering the optimal scrambling behaviour in the thermodynamic limit. Black dashed lines are for guiding the eye. Results are computed using exact diagonalisation (ED) for smaller system sizes, and a Krylov subspace method (see Methods~\ref{sec:krylov} for details) to reach larger system sizes. We see excellent agreement for the two when ED is possible. ED results are from $N=6$ to $N=13$, while Krylov results are up to $N=20$. Inset shows the average standard error in $\lambda$ from the Krylov subspace method vs $N$ which is larger for smaller system sizes.}
\label{fig:Lyapunov}
\end{figure}

One of the most important characteristics of black holes in a quantum context is that they are optimal scramblers \cite{Sekino_2008}. Moreover, this property is vital to the success of the Hayden-Preskill protocol \cite{yoshida2017}. While we have previously addressed this question using exact diagonalisation (ED), this severely limits the system sizes that can be investigated \cite{Daniel2024}. Therefore, we build upon these results and verify the optimal scrambling with a new numerical method that allows us to access larger system sizes.  We perform this investigation by considering the Lyapunov exponent $\lambda$, extracted from the regularised out-of-time-ordered correlators (OTOCs)~\cite{Shenker2014,Maldacena_2016,Romero_Berm_dez_2019}. To determine the Lyapunov exponent, $\lambda$, we fit an exponential-like function to the regularised OTOCs of the form
\begin{equation}\label{otocs}
    C(t)=\langle O_n(t)\rho^{1/4}O_m(0)\rho^{1/4}O_n(t)\rho^{1/4}O_m(0)\rho^{1/4} \rangle,
\end{equation}
where $\rho = \exp(-\beta H)/\mathcal{Z}$, with the partition function $\mathcal{Z}=\textup{Tr} \exp(-\beta H)$ and $\beta=1/T$ is the inverse temperature. 

In previous work~\cite{Daniel2024}, we observed that $\lambda$ appears to saturate the upper bound expected for quantum black holes, $\lambda=2\pi T(v/2)$ (here $v/2$ is the natural energy scale of the model when $|v|\gg |u|$), achieving optimal chaos as the system size $N$ increases. We fit a linear function of the form $\lambda = aT + c$ to $\lambda(T)$ obtained from the OTOCs utilising exact diagonalisation methods to determine if the scrambling is truly optimal. Due to this we were limited to system sizes in the interacting model of up to $N=13$. This produced significant finite size effects resulting to pronounced oscillations around the desired value for both $a$ and $c$, as shown in Fig.~\ref{fig:Lyapunov}(a)-(b). Here, we extend these results using an independent method available at~\cite{codegithub}, which uses a Krylov subspace method (see methods ~\ref{sec:krylov}). As shown in Fig.~\ref{fig:Lyapunov}(a)-(b), our new results are consistent with previous ED method, where we take $O_n=S_{N/2}^x$ and  $O_m=S_{N/2-2}^x$ for $N$ even and $O_n=S_{(N+1)/2}^x$, $O_m=S_{(N-3)/2}^x$, for $N$ odd in Eq.~\eqref{otocs}. Additionally, they go beyond the previous system sizes, reaching up to $N=20$, where the oscillations significantly subside. We find $a$ plateaus to the optimal value of $2\pi (v/2)$, while $c$ rapidly drops to 0, verifying our claim of optimal scrambling within the chiral spin-chain model.

\begin{figure}
    \centering \includegraphics[width=1\linewidth]{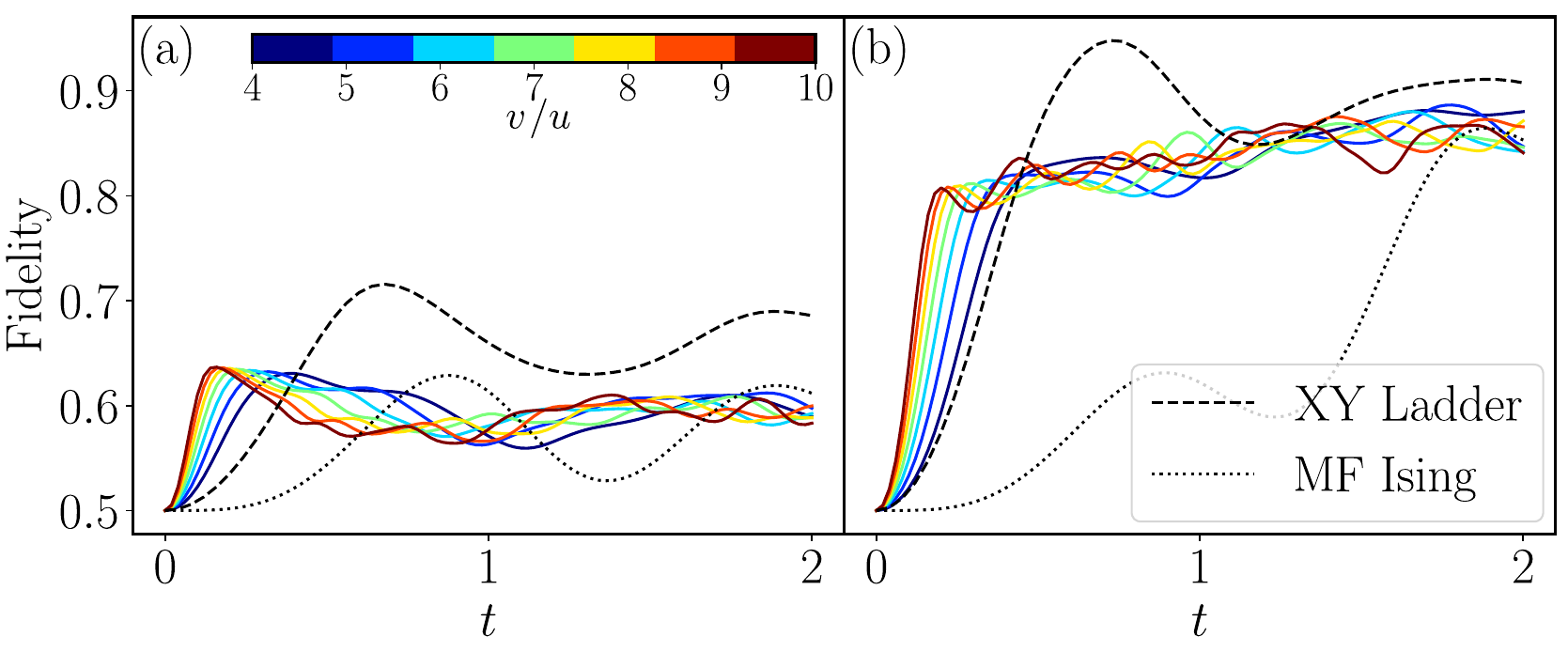}
    \caption{Comparison of teleportation fidelity over time after quenching the initial state with the chiral-spin model (solid lines), the XY ladder (dashed line), and the mixed field (MF) Ising model (solid dotted line). In the case of the mixed field Ising, the optimal ratio of values for chaos is used, as suggested in Ref.~\cite{Rodriguez2024}.(a) shows $E=1$, while (b) shows $E=4$ EPR measurements. Colour indicates the value of $v$ during the quench in the chiral-spin chain. Using comparable Hamiltonian parameters between the three models, we find the chiral-spin chain outperforms both the XY and mixed field Ising model, as evidenced by an earlier saturation in the teleportation success. The results are for $N=13$ using ED for all systems. For couplings, in the chiral model we take $u=-8$ such that the XY term has overall prefactor 1. For the XY ladder, we take the Hamiltonian from Ref.~\cite{Agarwal2023} with couplings $J_x$=0.91, $J_y=1$ while for MF Ising, we take the Hamiltonian from \cite{Rodriguez2024} with $J=1$, $h_x=1.1$ and $h_z$=0.3.}
    \label{fig:modelcompare}
\end{figure}

In principle, almost every many-body quantum Hamiltonian $H$ is chaotic, and should therefore scramble the system. In this sense almost every Hamiltonian is an equally valid choice to realise the Hayden-Preskill teleportation protocol ~\cite{Nakata2024}. Nevertheless, the chiral spin-chain model is a uniquely appropriate for two reasons. Firstly, the MFT description discussed above means the results can be understood in the large system limit as describing Dirac fermions in a curved spacetime. Secondly, the behaviour of OTOCs for the interacting system demonstrates that the system is optimally scrambling.  We now demonstrate the advantageous nature of the optimal scrambling found in the chiral-spin model for the teleportation procedure. In practicality, one would ideally complete the teleportation between Alice and Bob in the quickest time possible to minimise both experimental error, and to easily repeat the procedure. The shortest time possible comes from the time taken to maximally scramble the second term in Eq.~\eqref{eqn:telestate_mb}, which is determined by the scrambling time of the system. As the chiral spin-chain saturates the optimal bound for scrambling, demonstrated by the behaviour of the Lyapunov exponent shown in Fig.~\ref{fig:Lyapunov}(a)-(b), the teleportation protocol will become maximally efficient. 

We compared the teleportation protocol of the chiral-spin model to two other chaotic models, demonstrating that the chiral model exhibits more rapid teleportation. These models are the mixed-field Ising model \cite{Banuls2011,Kim2013,Rodriguez2024}, a commonly studied chaotic model, and the 2D ladder XY model where the same teleportation protocol was studied in Ref.~\cite{Agarwal2023}. For the former, we use the model parameters identified in Ref.~\cite{Rodriguez2024} where these were found to correspond to a maximum entanglement distribution, corresponding to its maximally chaotic point. We scale the chiral model parameters such that the energy scales of the three models are comparable and compare for different values of $v/u$. 
We see in Fig.~\ref{fig:modelcompare}(a)-(b) that the chiral model outperforms both models for $v$ larger than the phase transition, demonstrating the optimal scrambling in effect and the advantage of this model.

Finally, we estimate the scrambling time $t_{\text{scr}}$, which quantifies how quickly the black hole interior thoroughly mixes quantum information. It is believed $t_{\text{scr}}$ to be inversely proportional to the Lyapunov exponent $\lambda$ as $
t_{\text{scr}} \propto \frac{1}{\lambda}$.
While for small temperatures, the chiral spin-chain model saturates the bound on chaos
$\lambda/(v/2) = 2\pi T $, during the teleportation protocol the system is effectively infinite temperature due to its maximal entanglement with the outside. Our numerical results (see Supplementary material) indicate that at that regime the Lyapunov exponent is approximately $\lambda \approx 0.78 v$.
Thus, the scrambling time follows
$t_{\text{scr}} \propto {1}/{v}$,
which is consistent with our teleportation results in Fig.~\ref{fig:chiralteleportation}(a).

\subsection{Butterfly Velocity and the propagation of information}
\label{sec:butterflyvelocity}

\begin{figure}
    \centering
    \includegraphics[width=1\linewidth]{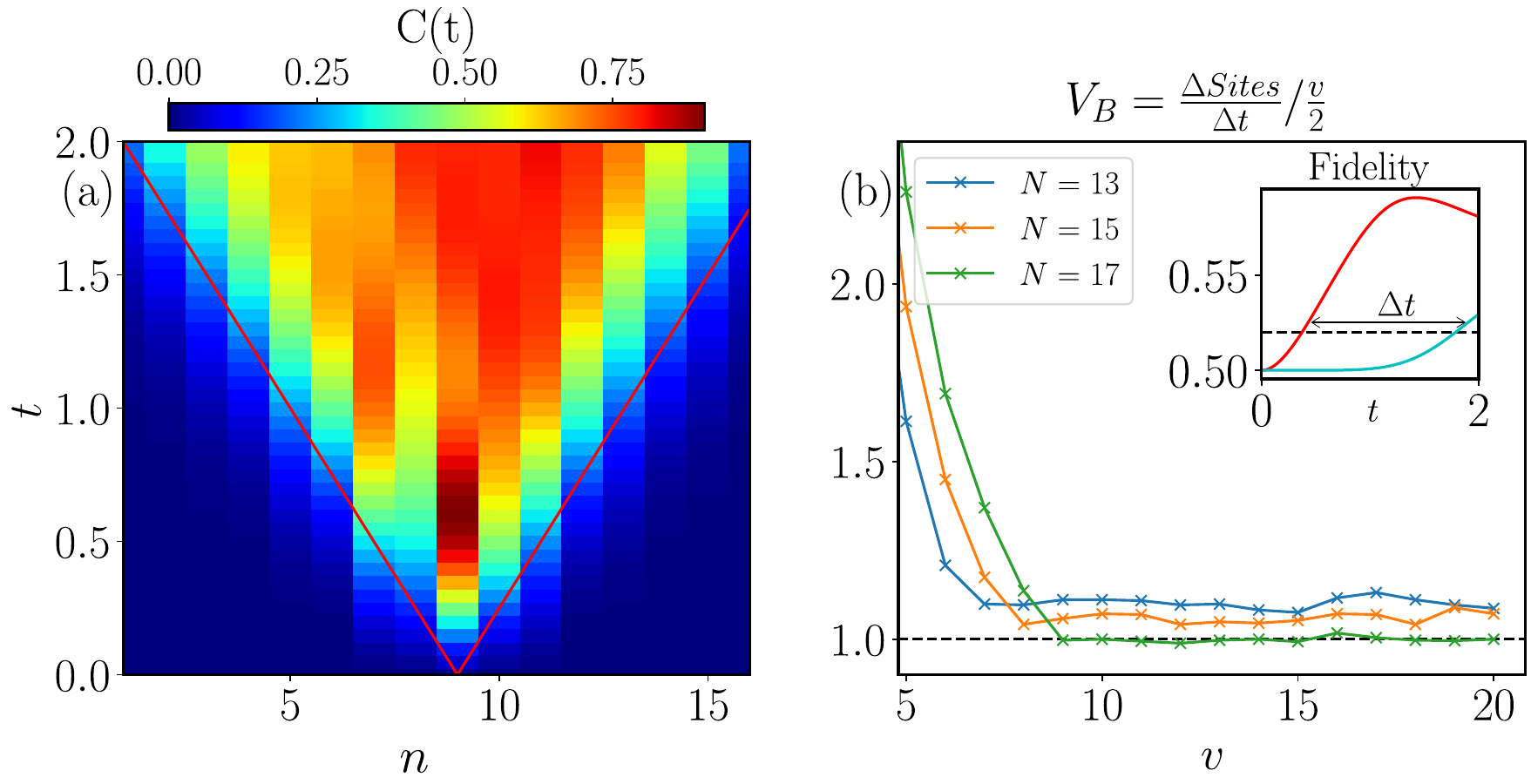}
\caption{Butterfly velocity and black hole teleportation time.
(a) is the Out-of-time order correlators (colour scale) over time measured on site $i$. A linear fit is then applied to estimate the butterfly velocity which is found to be $V_B\approx v/2$. Results are computed using a Krylov algorithm for $N=17$, $v=8$, $O_n=\sigma^Z_n$ at infinite temperature in Eq.~\eqref{otocs}. (b) We estimate the speed of propagation of quantum information by comparing the fidelity when measuring the outer EPR pair first versus the inner EPR pair first during the projection in the teleportation protocol shown in Fig. \ref{fig:sketch}(a). We find the time difference, $\Delta t$, for when these two fidelities match (see Inset), indicating the time spent for information to propagate between these two EPR pairs positioned at distance $\Delta l$. We use this to estimate the butterfly velocity, $V_B=\Delta l/\Delta t$, which remarkably saturates to $V_B=v/2$ with increasing system size.}
\label{fig:butterflyvelocity}
\end{figure}

The speed of the teleportation protocol also depends upon the order the EPR measurements are performed, due to the finite time necessary for information to propagate along the chain between the initial teleportation qubit and the position of the first EPR measurement. The black hole Hamiltonian, $H$, will immediately begin scrambling the EPR state on the site where the $\sigma^a_A$ rotation is acting (cf. Eq.~\eqref{eqn:telestate_mb}). This is the only site where the state is different from the $\ket{EPR}$ state, which is an eigenstate of the full scrambling Hamiltonian $H$. Therefore, for EPR pairs far from this state to become scrambled, the effect needs to propagate there with speed set by the butterfly velocity, $V_B$. For example, measuring the inner EPR pair in contrast to the outer (see Fig.~\ref{fig:butterflyvelocity}(a)), will cause a delay in the teleportation success due to the need for information to first propagate across their respective locations before scrambling can take place~\cite{Keselman2021}. 

We calculated the delay created due to the measurement order in the following manner. Firstly, we measured the gradient of the regularised OTOCs, as in Eq.~\eqref{otocs}, along the chain - a typical method used to estimate the butterfly velocity. The results are shown in Fig.~\ref{fig:butterflyvelocity}(a) where the gradient of the red line is found to be $V_B\approx v/2$. We note the difference in the left and right propagation is due to the chirality of the model which inversion symmetry. 
Secondly, we compared the fidelity when projectively measuring on different sites, and calculate the time required for the teleportation fidelity to reach matching values, $\Delta t$. Dividing the difference in lattice sites of the two measurements, $\Delta l$, by the time lapse $\Delta t$, we independently obtain the value of the butterfly velocity. In practice we compared the fidelity when projectively measuring the EPR pair on the first site to the $(N-1)/2$th site such that $\Delta l = (N-1)/2 - 1$. Scanning this for different $v$ in Fig.~\ref{fig:butterflyvelocity}(b) we similarly find $V_B$ saturates to approximately $v/2$ deep into the chiral phase. Therefore, the measurement order of EPR pairs affects the speed of the teleportation protocol due to scrambling propagating through the system with butterfly velocity.
This propagation time is given by
$t_B = {\Delta l}/{V_B}\propto 1/v$. This differs from an effective model of a black hole such as the SYK model, which is defined by randomised all-to-all coupling, and therefore lacks a well defined notion of distance. 

\subsection{Krylov subspace method}
\label{sec:krylov}
In this article we have employed several numerical techniques, the most notable one being a matrix-free Kyrlov subspace method used to achieve higher system sizes than would be possible with the standard exact diagonalisation method. More specifically, we do not calculate Eq.~\ref{otocs} exactly, but instead compute the following quantity
\begin{multline}
\tilde C_{\ket{\Phi}}(t)=\bra{\Phi} e^{-\beta H/8 + itH}O_le^{-\beta H/4 - itH}O_k\\e^{-\beta H/4 + itH}O_le^{-\beta H/4 - itH}O_ke^{-\beta H/8}\ket{\Phi}
\end{multline}\label{krylov otocs}%
where $\ket{\Phi}$ is a (normalised) Haar-random state whose components are generated from a Gaussian distribution with zero mean and unit variance. The true value of the OTOC $C(t)$ is then approximated by averaging $\tilde C_{\ket{\Phi}}(t)$ over several thousands of different samples of $\ket{\Phi}$. With a matrix-free representation of the Hamiltonian $H$, which we explain in detail below, this method allows us to eliminate the memory overhead of storing the full matrix form of the Hamiltonian, achieving a dramatic reduction of the memory requirement to that of a small number of Hilbert space vectors. The computation is based on an open-source library \cite{gpulib} that implements this Krylov subspace method efficiently on the GPU, and the source code is publicly available at \cite{codegithub}.

To understand this reduction of the memory requirement, one notes that since in general the Hamiltonian is written as a sum of the product of Pauli operators
\[
H = \sum_{i_1,\ldots,i_N} \lambda_{i_0 \ldots i_N}\,
\sigma_{i_0}\cdots\sigma_{i_N}
\]
where $\lambda_{i_0\cdots i_N}$ are complex numbers and each $\sigma_i$ is one of $\{I, \sigma^x,\sigma^y,\sigma^z\}$ defined on site $i$, it is unnecessary to store the full $2^N$ by $2^N$ dimensional matrix form of the Hamiltonian in memory. Instead, storing the coefficients $\lambda_{i_0 \ldots i_N}$ is sufficient to compute the action of the Hamiltonian on a state, since the actions of each Pauli operators on the state can be trivially realised with (a combination of) swapping and scalar multiplications on the coefficients of the state vector. Starting with a state vector $|v\rangle$, using this this matrix-free action of the Hamiltonian one can build a sequence of vectors
\[
|v\rangle, H|v\rangle, H^2|v\rangle,\cdots, H^n|v\rangle
\]
which spans a linear subspace, known as a Krylov subspace of the full Hilbert space. Diagonalizing within the Krylov subspace and iterating this process then allow one to compute the action of the exponential of the Hamiltonian on the state vector to an arbitrary degree of precision. The GPU-computing library \cite{gpulib} that we have employed in this project implements the restarted Krylov subspace method as described in Ref.~\cite{eiermann2006}. We choose the dimension of the Krylov subspace $n=5$ which we have found to strike a balance between computational speed and memory usage, and set the tolerance of the calculation (given by the $L_2$-norm of the difference of the result vectors between two iterations) to the machine precision.

Contrasting with the standard exact diagonalisation method commonly used in computing the matrix exponential, we have found that the Krylov subspace method leads to considerably faster computational time and much reduced memory requirement, while offering the same accuracy, as is shown in the main text where we have compared the results of either methods in the regime where both can be applied. The reduction of the storage of a $2^N$ by $2^N$ matrix to a few $2^N$-dimensional vectors allows the Krylov subspace method to reach twice the system size than is feasible with the exact diagonalisation method, given the same amount of computational resources. In general, the convergence of the Krylov subspace method depends on the norm of the operator being exponentiated, and we have found that the method takes longer time to converge as one increases $N$, $\beta$, and $t$. It is expected that the method eventually breaks down for sufficiently large system size, and our code will throw errors in this case so the user will be informed that the calculation is no longer to be trusted.

\section{Conclusions and Discussion}\label{sec:conclusion}

In this work, we demonstrate a realisation of the Hayden-Preskill quantum teleportation protocol within a binary black hole system simulated by a chiral spin-chain. This model successfully demonstrates the key ingredients of the protocol: the generation of entanglement through Hawking radiation and the implementation of optimal scrambling dynamics within the black holes. The optimal scrambling facilitates rapid information dispersal, reducing the overall time required for teleportation. By using these properties of the chiral spin-chain, we have shown that it is possible to achieve high-fidelity quantum teleportation even with relatively small system sizes. We further quantified the propagation speed of quantum information using the butterfly velocity, finding that it aligns well with theoretical expectations and saturates to $v/2$ in the strongly interacting regime. The scrambling time and the butterfly velocity of the chiral spin-chain underscore the efficiency of the black hole teleportation protocol. All the characteristic timescales involved in black hole teleportation scale inversely with $v$, the coupling strength of the chiral spin-chain model.

Comparative analyses with other chaotic models, such as the mixed-field Ising model and the XY ladder, revealed that the chiral spin-chain outperforms them in terms of scrambling speed. This minimises the time required for information recovery and thus increases the efficiency of the quantum teleportation protocol. Our black hole simulator no only can achieve optimal scrambling, but it can also support a mean-field geometric regime that faithfully reproduces semiclassical black hole features such as Hawking radiation and the Page curve. In contrast to previous models that simulate either geometric backgrounds~\cite{horner2023} or scrambling behaviour in isolation~\cite{Daniel2024}, the chiral spin-chain provides a unified platform that captures both aspects simultaneously. Hence, it enable us to reproduce key features of black holes, such Hawking radiation, the Page curve, butterfly velocity scaling, and optimal scrambling, within a single consistent framework. This synergy is essential for realising the full structure of the Hayden-Preskill teleportation protocol and for drawing robust analogies with quantum gravitational dynamics. 

Future work may focus on extending these results to higher-dimensional systems, and investigating the interplay between scrambling dynamics and other quantum error correction protocols in holographic settings. Additionally, while we have offered strong numerical evidence of this optimal scrambling, an analytic proof remains elusive. We hope to further analyse the low energy, effective physics of the model using a field theory approach such a bosonisation, to derive the optimal scrambling result we have demonstrated numerically. 

Finally, we will explore an algorithmic circuit realisation of the black hole scrambler as a practical route to experimentally implement our protocol. In particular, by Trotterising the chiral spin-chain Hamiltonian, we can decompose its time evolution into a sequence of local gate layers that approximate the scrambling dynamics through repeated application of simpler interactions. Specifically, the two-body XX interactions and the three-body chiral terms can each be expressed using controlled rotations and two qubit Givens rotations, which are widely supported on current quantum hardware. Our initial analysis shows that to faithfully capture the scrambling dynamics and achieve effective teleportation fidelity, approximately 5 Trotter rounds are sufficient. For each round, the gate count scales linearly with the system size. So for a system of $N \approx 12$ spins, this corresponds to a total of around several hundred gates, a regime within reach of present-day quantum processors and programmable cold-atom platforms~\cite{benhemou2025probingquantumpropertiesblack}. By using this Trotterised approach, the full Hayden-Preskill teleportation sequence that includes scrambling, EPR pair preparation, and measurement, can be implemented with standard quantum algorithm techniques. Moreover, this digital framework allows direct probing of scrambling dynamics through measurable quantities such as out-of-time-ordered correlators and Loschmidt echoes, offering a concrete and practical pathway to realise black hole information processing in laboratory conditions.
\newline
\section{Acknowledgements}
We would like to thank Matthew Horner, Iason Sofos and Cristian Voinea for inspiring conversations. This work was supported by EPSRC with Grant No. EP/Z533634/1 and UKRI1337: Anyons24 and the Leverhulme Trust Research Leadership Award RL-2019-015.
\bibliography{ref}

\clearpage


\newpage

\appendix

\title{Supplementary Information for ``Quantum teleportation between simulated binary black holes''}

\maketitle
\date{\today}


\section{Numerical scaling}\label{sec:Numerical}

In this section, we present relevant numerics data on the dependence of the teleportation scheme on system size and on the high temperature behaviour of the Lyapunov exponent for the chiral spin-chain model.

\subsection{System size scaling}\label{sec:Ncompare}
\begin{figure}
    \centering
    \includegraphics[width=1\linewidth]{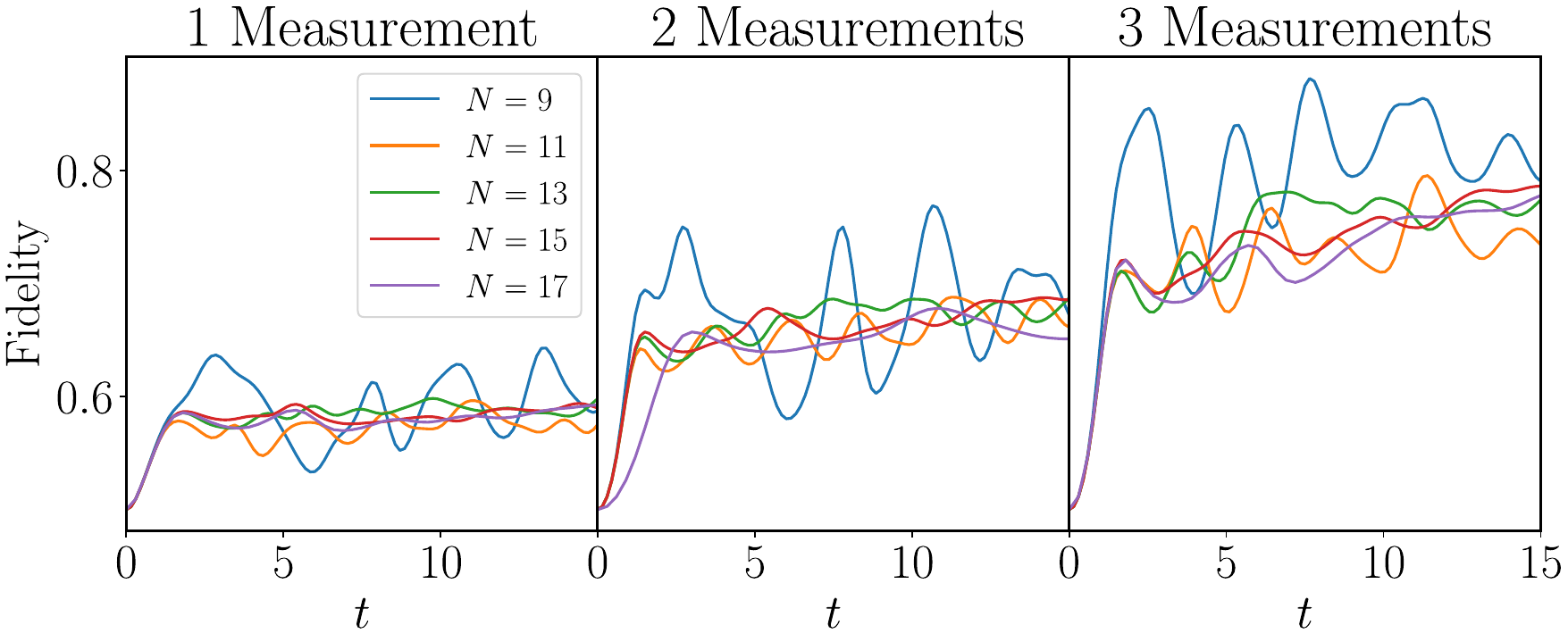}
    \caption{A comparison of the Fidelity of teleportation with system size $N$ using ED for $u=1$, $v=8$. Three panels show the 1, 2 and 3 simultaneous epr measurements respectively. We see with increasing system size, scrambling speed remains the same while the fidelity plateau becomes increasingly evident.}
    \label{fig:Ncompare}
\end{figure}

In this subsection we present results with varying system size to demonstrate the robustness of the protocol with $N$, expanding upon the results of Fig. 3(a) in the main text. In Fig.~\ref{fig:Ncompare}, we see that as we increase system size, the fidelity does not drop, but instead increasingly plateaus at a constant value while the scrambling speed remains constant. This indicates a level of robustness with system size, and suggests this will fidelity plateau will persist in the large $N$ limit.

\subsection{Lyapunov exponent in the infinite $T$ limit}\label{infiniteNlyap}
\begin{figure}
    \centering
    \includegraphics[width=1\linewidth]{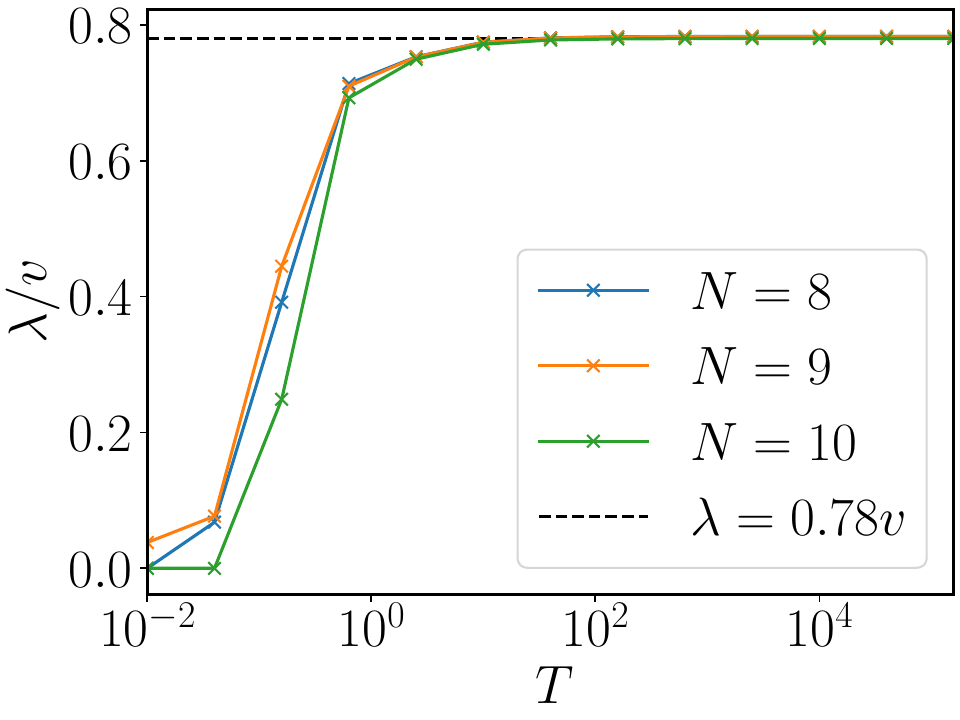}
    \caption{The scaled Lyapunov exponent, $\lambda /v$, versus temperature, $T$, where colour indicates system size $N$. We see in the large $T$ limit, $\lambda/v$ rapidly saturates to the value $\lambda=0.78v$ with little change with system size. Results are computed using ED.}
    \label{fig:lyapinfiniteN}
\end{figure}

In Section (A) of methods, we stated that in the infinite temperature limit, the Lyapunov exponent saturates to a value $\lambda=0.78v$ in the Chiral model. Here we supplement that with numerical verification. Fig.~\ref{fig:lyapinfiniteN} demonstrates this where we see a rapid saturation to $0.78v$ with increasing  $T$ irrelevant of system size.

\section{Population dynamics and Page times} \label{sec:Popdynamics}

\begin{figure}
\includegraphics[width=1\linewidth]{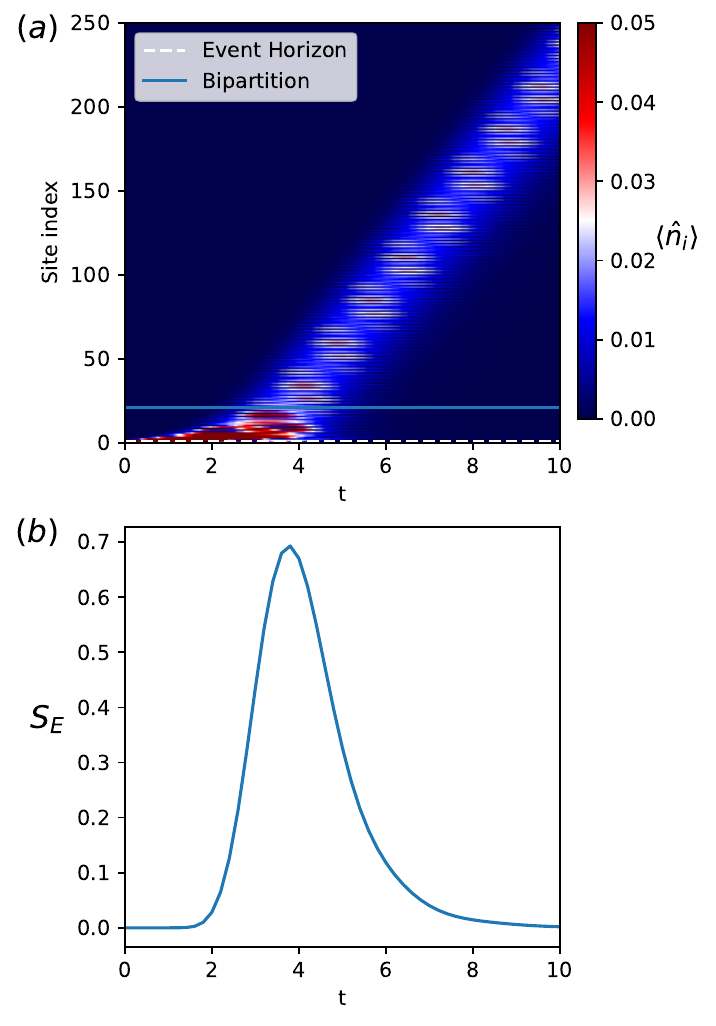}
\caption{ (a) On-site population dynamics $\langle \hat{n}_i \rangle = \langle c_i^{\dagger}c_i\rangle$ and (b) Page curve for the MFT Hamiltonian. Results are shown for $N = 500$ with 1 particle initialised inside the black hole with the horizon at the second site and $(\alpha, \beta) = (15,0.05).$}
\label{fig:pop_dynamics}
\end{figure}

\begin{figure}
\includegraphics[width=1\linewidth]{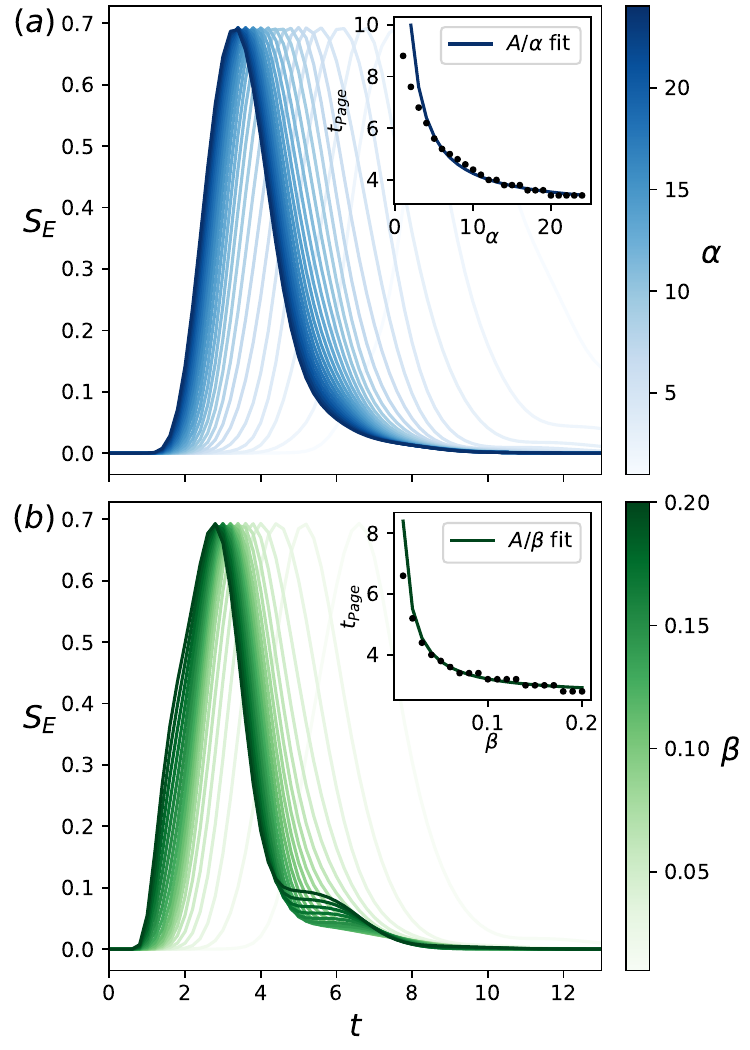}
\caption{ (a)-(b) Page curves and Page times (inset) as a function of $\alpha, \beta$ in the MFT Hamiltonian. Results are shown for $N = 500$ with 1 particle initialised inside the black hole with the horizon at the second site.}
\label{fig:page_times_fit}
\end{figure}

In this section, we calculate the population dynamics of the MFT model presented in the main text to give a physical picture of the outgoing Hawking radiation. At first, we consider the MFT Hamiltonian in Eq. (1) with the couplings set by Eq. (6). We initialise a single particle inside the horizon which is set at the second site. Then, we monitor the on-site population dynamics $\langle \hat{c}_i^\dagger \hat{c}_i\rangle$ and calculate the entanglement entropy for a bipartition of the chain at $n = 22$. The partitioning site is chosen away from the horizon to avoid boundary effects arising from reflections from the end of the chain. In Fig. \ref{fig:pop_dynamics}(a), we observe that the particle population, initially localised at the first site, eventually leaks out of the horizon. This is accompanied by a linear growth in the entanglement entropy in Fig. \ref{fig:pop_dynamics}(b), which peaks to $\log(2)$ at the Page time when half of the particle population escapes the partition site. Successive smaller peaks are seen in the entanglement entropy which result from reflections of the particle from the boundary of the chain, these match the population spikes seen in Fig. \ref{fig:pop_dynamics}(a).

We also consider the Page curves as a function of $\alpha$ and $\beta$ in Fig. \ref{fig:page_times_fit}(a)-(b). Considering a similar setup as in Fig. \ref{fig:pop_dynamics}(a)-(b), the Page times decrease monotonically upon increasing $\alpha, \beta$. This is a consequence of the Hawking radiation physics: $t_{Page} \propto 1/T_H$, and the Hawking temperature $T_H$ is proportional to both $\alpha$ and $\beta$. Indeed, we observe a $1/\alpha$ and a $1/\beta$ dependence of the Page times in the insets of Fig. \ref{fig:page_times_fit}(a)-(b).

\end{document}